\documentclass[journal]{IEEEtran}
\IEEEoverridecommandlockouts
\usepackage{cite}
\usepackage{amsmath,amssymb,amsfonts}
\usepackage{algorithmic}
\usepackage{graphicx}
\usepackage{textcomp}
\usepackage{xcolor}
\usepackage{gensymb}
\usepackage{makecell}

\begin{document}
%
\title{Low-Loss Silicon Directional Coupler with Arbitrary Coupling Ratios for Broadband Wavelength Operation Based on Bent Waveguides
  
\thanks{This work was supported by imec's industry-affiliation R\&D program “Optical I/O”.}%
}


\author{
Ahmed H. El-Saeed,
Alaa Elshazly,
Hakim Kobbi,
Rafal Magdziak,
Guy Lepage,
Chiara Marchese,
Javad Rahimi Vaskasi,
Swetanshu Bipul,
Dieter Bode,
Marko Ersek Filipcic,
Dimitrios Velenis,
Maumita Chakrabarti,
Peter De Heyn,
Peter Verheyen,
Philippe Absil,
Filippo Ferraro,
Yoojin Ban,
Joris Van Campenhout,
Wim Bogaerts,
and Qingzhong Deng
\thanks{A. H. El-Saeed and A. Elshazly are with imec, Kapeldreef 75, 3001 Leuven, Belgium and the Photonics Research Group, Department of Information Technology, Ghent University-imec, Ghent, Belgium (email: ahmed.bayoumi@imec.be; alaa.elshazly@imec.be)}
\thanks{H. Kobbi, R. Magdziak, G. Lepage, C. Marchese, J. R. Vaskasi, S. Bipul, D. Bode, M. E. Filipcic, D. Velenis, M. Chakrabarti, P. D. Heyn, P. Verheyen, P. Absil, F. Ferraro. Y. Ban, J.V. Campenhout, and Q. Deng are with imec, Kapeldreef 75, 3001 Leuven,
Belgium (email: hakim.kobbi@imec.be; rafal.magdziak@imec.be; guy.lepage@imec.be; chiara.marchese@imec.be; javad.rahimivaskasi@imec.be; swetanshu.bipul@imec.be; dieter.bode@imec.be; marko.filipcic@imec.be; dimitrios.velenis@imec.be; maumita.chakrabarti@imec.be; peter.deheyn@imec.be; peter.verheyen@imec.be; philippe.absil@imec.be; filippo.ferraro@imec.be; yoojin.ban@imec.be; joris.vancampenhout@imec.be; qingzhong.deng@imec.be)}
\thanks{W. Bogaerts is  with the Photonics Research Group, Department of Information Technology, Ghent University-imec, Ghent, Belgium and IMEC, Kapeldreef 75, 3001 Leuven, Belgium (email: wim.bogaerts@ugent.be)}
\thanks{Corresponding authors: Q. Deng and W. Bogaerts}

}

\maketitle

\begin{abstract}

  We demonstrate a design for a high-performance $2 \times 2$ splitter meeting the essential requirements of broadband coupling, support for arbitrary coupling ratio,  ultra low-loss, high fabrication tolerance, and a compact footprint.
  This is achieved based on a rigorous coupled mode theory analysis of the broadband response of the bent directional coupler (DC) and by demonstrating a full coupling model, with measured broadband values of 0.4, 0.5, 0.6, and 0.7.
  As a benchmark, we demonstrate a 0.5:0.5 splitter that significantly reduces coupling variation from 0.391 in the traditional DC to just 0.051 over an 80 nm wavelength span.
  This represents a remarkable 7.67 times reduction in coupling variation.
  Further, newly-invented low-loss bends were used in the proposed design leading to an ultra low-loss design with negligible excess loss ($\mathrm{0.003 \pm 0.013 \ dB}$).
  The proposed 0.5:0.5 silicon strip waveguide-based design is tolerant and shows consistently low  coupling variation over a full 300 mm wafer showcasing a maximum cross coupling variation of 0.112 over 80 nm wavelength range, at the extreme edge of the wafer.
  Futhermore, we augmented the wafer mapping with a waveguide width fabrication tolerance study, confirming the tolerance of the device with a mere 0.061 maximum coupling variation with a waveguide width deviation of $\pm 20$ nm over 80 nm wavelength range.
  These specs make the proposed splitter an attractive component for practical applications with mass production.
  
  \end{abstract}



  \section{Introduction}

Within the realm of photonic integrated circuits (PICs), optical power splitters serve as  indispensable components. Among these, the $2 \times 2$ splitters emerge as fundamental building blocks with significant applications such as modulation~\cite{HybridSiliconLithium_IJOSTIQE2021a}, signal switching~\cite{LowLossBroadband_OL2016a}, and wavelength division multiplexing (WDM)~\cite{32x100GhzWdm_2023a}.
  In particular it is often essential in those applications to design broadband $2 \times 2$ splitters capable of both balanced and unbalanced splitting.
  Furthermore, since a large number of these devices are often used in PICs, 
  they must exhibit minimal excess loss, maintain a compact footprint, and not complicate the fabrication process.

The directional coupler (DC) has traditionally served as the primary $2 \times 2$ splitter owing to its simplicity and ease of fabrication.
  The DC operates through wavelength-dependent evanescent coupling between two closely spaced waveguides, allowing the oscillation of light between these waveguides. 
  This ultimately enables the splitting of incident light into the through and cross ports. 
  Although DCs are theoretically lossless, support arbitrary splitting ratios, and possess a compact design, their performance undergoes significant degradation owing to their inherent wavelength dependence due to dispersion.
  For instance, a simple silicon DC designed for $0.5:0.5$ splitting has more than $0.53$ coupling variation over an 80 nm wavelength span~\cite{TandemMachZehnder_S2020a}. 

In the pursuit for an arbitrary-splitting-ratio, broadband, low-loss, robust, and compact splitter, various design schemes have been investigated in the literature.
Multi mode interferometers (MMIs) are commonly used as broadband $2 \times 2$ splitters, where the wavelength dependence is averaged out through the different dispersion profiles of the high order modes propagating through the MMI~\cite{OpticalMultiMode_JOLT1995b}.
However, MMIs only allow for a few discrete coupling ratio choices, and more complex configurations are a must to enable arbitrary splitting ratios~\cite{BroadbandAngledArbitrary_JOLT2020a}.
Further, MMIs often suffer from high excess loss and output imbalance, and are quite sensitive to the position of the input and output ports~\cite{DesignExperimentalAnalysis_OE2020a}.
Adiabatic DCs (ADCs) are another common approach to achieve broadband coupling with arbitrary coupling ratios~\cite{UltraBroadbandChip_OE2024a, CompactUltraBroadband_IPTL2024a, AdiabaticCouplerDesign_JOLT2019a}. 
Unlike DCs, only one mode is excited in adiabatic couplers, and this mode evolutes adiabatically through the coupler and therefore broadband coupling could take place~\cite{AdiabaticityEngineeredSilicon_IPJ2023a}.
ADCs are, however, inherently long devices that could be longer than  240 $\mathrm{\mu m}$ and often exhibit high excess loss~\cite{AdiabaticCouplerDesign_JOLT2019a, SiliconInsulatorBased_OL2013a}.

On the other hand, several attempts have been made to adapt the traditional DC for broadband operation.
For instance, DCs based on rib waveguides have shown broadband coupling through dispersion engineering of the geometrical parameters. 
However, their input and output ports require large bending radii, as large as 180 $\mathrm{\mu m}$, in order to minimize the bending loss, resulting in a large footprint~\cite{WavelengthIndependentDirectional_JOLT2017a}. 
Another approach is based on incorporating phase-compensation sections to the DC design. 
However, despite their broadband coupling, these designs are sensitive to linewidth variations and further design considerations are a must to make them fabrication tolerant~\cite{BroadbandSiliconPhotonic_OE2015a, ComparisonFabricationTolerance_2019a}. 
Another strategy is to cascade DCs in a  Mach-Zehnder interferometer (MZI) configuration, where an optical path length difference is introduced to compensate for the wavelength dependence of the DC.
Arbitrary coupling ratios are possible via the optimization of the DC coupling ratios and the MZI arms lengths~\cite{FabricationTolerantWavelength_OE2022a, TolerantBroadbandTunable_OE2020a}. 
However, precise control over the DC coupling ratios and the phase difference of the MZI arms is a necessity and ultimately complicates the design process. 
Further, subwavelength gratings (SWGs) have shown the capability to render the DC broadband by means of fine tuning the periodic SWG structures incorporated into the DC~\cite{UltraBroadbandCompact_P2023a}. 
However, the fabrication of the SWGs is complex and unreliable specifically with feature sizes as small as 70 nm~\cite{UltraCompactSubwavelength_IPTL2019a, UltraBroadband2_OL2018a}.
Finally, by bending the waveguides in the DC, a seven-fold enhancement in the operational bandwidth of the $0.5:0.5$ coupler was possible as compared to the traditional straight DC, all while maintaining a compact footprint, being fabrication tolerant and insensitive to temperature variations~\cite{BendingPlanarLightwave_IPTL2005a, ReductionWavelengthDependence_JOLT2014a}. It should be noted that with the introduced asymmetry, the $\pi/2$ phase difference between the through and cross ports is not necessarily satisfied anymore.
In bent DCs, the introduction of asymmetry, through bending waveguides with different bending radii, eliminates the need for different waveguide widths.
This eventually addresses the fabrication sensitivity observed in DCs with different waveguide widths~\cite{BendingPlanarLightwave_IPTL2005a}. 
Several studies employed the bent DC to achieve broadband coupling with different approaches, including using the transfer matrix method to solve for the design parameters  \cite{BroadbandSiliconInsulator_SR2017a}, using a cascade of bent DCs to achieve polarization insensitive and broadband coupling response ~\cite{PolarizationInsensitiveBroadband_OL2017a}, and using a semi-inverse design method to optimize freely-shaped waveguide sections shapes~\cite{HighPerformance2×2_JOLT2023a}. 
Nonetheless, a detailed analysis of the bent DC wavelength dependence and an experimental-based coupling model for achieving broadband coupling at arbitrary coupling ratios have not yet been developed.

\begin{figure}[!b]
  \centerline{\includegraphics{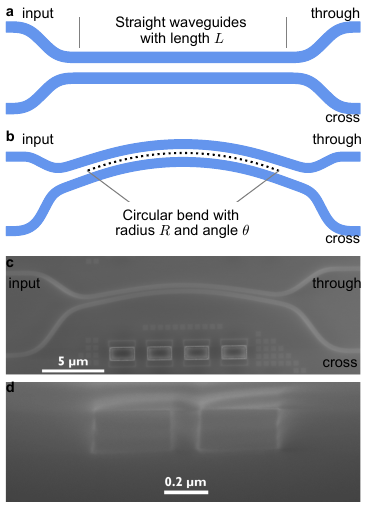}}
  \caption{Schematic of the traditional straight DC with $L$ as the coupling length (a).
  Schematic (b) and scanning electron microscope (SEM) image (c) of the proposed bent DC with $R$ as the coupling radius, and $\theta$ as the coupling angle.  All bent DC curves are designed with low-loss bends~\cite{LowLossWaveguide_2023a}, while the straight DC input and output ports have traditional circular bends with a bending radius of $\mathrm{5\ \mu m}$. SEM image of the cross section of the proposed bent DC in the coupling region (d).   The waveguide material stacks are SOI with silicon oxide as top cladding, using IMEC iSiPP300 platform. All DCs are based on strip waveguides with nominal silicon thickness of 220 nm, and waveguide width of 380 nm. The coupling gap is nominally 100 nm. 
  \label{schmatic}}
\end{figure}


\begin{figure*}[!h]
  \centerline{\includegraphics{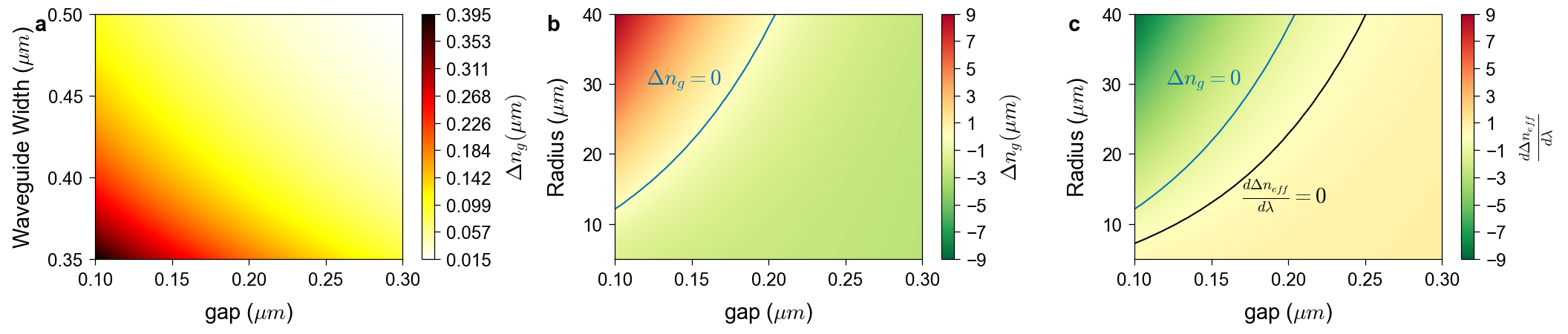}}
  \caption{Wavelength dependence analysis for the straight and bent DC with lines showing broadband criteria. Contour plot of $\Delta n_g $ as a function of the gap and the waveguide width for the straight symmetric DC (a). Contour plots of  $\Delta n_g $ (b) and $d\Delta n_{eff} / d \lambda $ (c)  for the bent DC as a function of gap and radius, where the waveguide widths are fixed as $\mathrm{0.38\ \mu m}$, along with lines showing where $\Delta n_g = 0 $ (blue) and $d\Delta n_{eff} / d\lambda = 0$ (black).
  \label{bdc_contour_plots}}
\end{figure*}

In this work, an analytical model describing the broadband behavior of the bent DCs is derived by means of a rigorous coupled mode theory (CMT) analysis.
This model can be used to design high-performance broadband bent DCs with arbitrary coupling ratios as experimentally verified in this paper.
We have already presented some initial results of the bent DC design~\cite{BroadbandBentDC_IEEEConf}, while the thorough analysis to derive broadband behavior of the bent DC and the experimental model to design broadband bent DCs at arbitrary coupling ratios will be discussed here. 

To benchmark with the literature, a 0.5:0.5 bent DC is demonstrated with a compact length of 27.5 $\mathrm{\mu m}$ and a coupling variation of 0.051 over 80 nm wavelength range which is the least coupling variation as compared to the reported bent DCs to our best knowledge.
Moreover, low-loss bends~\cite{LowLossWaveguide_2023a} that enable continuous curvature and curvature derivative at all connections are introduced into the 
bent DC, resulting in the lowest coupling loss ($\mathrm{0.003 \pm 0.013 \ dB}$) among all the silicon $2 \times 2$ splitters.
Last but not least, the proposed DCs are robust and could be reliably used in mass production as indicated by wafer scale measurements on imec's 300 mm platform.

\section{Coupling wavelength dependence analysis for straight and bent DC}

\begin{figure}[h]
  \centerline{\includegraphics{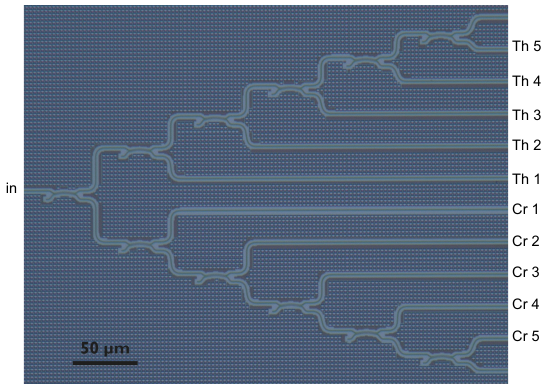}}
  \caption{Microscope image of the cascaded identical bent DCs used for robust measurements.  The power is input at the port (in) and measured at all the cascaded output stages. The labels determine whether the cascaded through (Th) or cascaded cross (Cr) coupled power is extracted, and the number indicates the number of times the measured value (through or cross) was repeated for that measurement. The bent DC has a bending radius of $\mathrm{25\ \mu m}$, a gap of $\mathrm{0.1\ \mu m}$,  and a coupling angle of $\mathrm{8.5\ degrees}$.
  \label{cascade_bent_dc}}
\end{figure}

\begin{figure}[!b]
  \centerline{\includegraphics{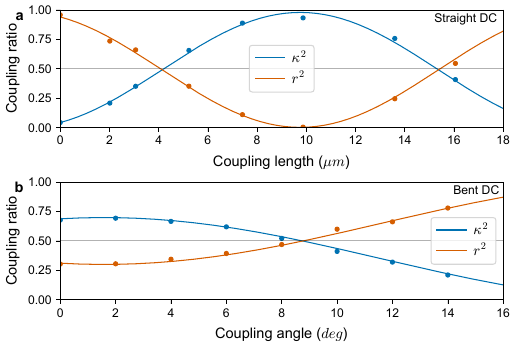}}
  \caption{Sinusoidal fitting of the cross coupling ($\kappa^2$) and  through coupling ($r^2$) with the coupling length for the straight DC (a) 
  and with the coupling angle for the bent DC (b) at $\mathrm{ \lambda = 1.31 \ \mu m}$ for the proposed devices.
  Both desings have a  gap of $\mathrm{0.1\ \mu m}$.  
  The bent DC has a bending radius of $\mathrm{25\ \mu m}$.
  Dots represent measured values, and lines depict the fitting.
  \label{sin_fitting}}
\end{figure}

According to CMT \cite{PhotonicDevices_2005a}, the through ($r^2$) and cross ($ \kappa^2$) coupling ratio of a directional coupler (Fig.~\ref{schmatic}) can be expressed as:
\begin{equation}
  \begin{cases}
  r^2=1-m \sin^{2}(\beta_c l+\phi), \\
  \kappa^2=m \sin^{2}(\beta_c l+\phi),
  \label{eq_for_pc}
  \end{cases}
\end{equation}

where the excess loss is ignored, $m$ is the matching coefficient and determines the maximum coupling ratio which is inversely correlated with the asymmetry, i.e. $m < 1$ for a  DC with non-negligible asymmetry, and $m = 1$ for a symmetric straight DC.
$\beta_c = \pi (n_{eff, even} - n_{eff, odd})/ \lambda$ denotes the coupling strength per unit length or angle, where $n_{eff,even}$ and $n_{eff,odd}$ denote the effective refractive indices of the even or odd supermodes in the DC, respectively. 
Further,  $l=L$ is the coupling length in a straight DC (Fig.~\ref{schmatic}a),
and $l=\theta$ is the coupling angle in a bent DC (Fig.~\ref{schmatic}b, c).
Additionally, $\phi$ accounts for the coupling contributed from the input and output connection bends.
Note that the effective refractive indices related to bent waveguides should be calculated with Maxwell's equations expressed in a cylindrical coordinate system.

The wavelength dependence for the cross coupling of a directional coupler can be calculated by taking the derivative of $\kappa^2$ (Eq.~\ref{eq_for_pc}), as shown: 

\begin{equation}
  \begin{split}
    \frac{d\kappa^2}{d \lambda} &= \frac{d m}{d \lambda} \sin^{2}(\beta_c l+\phi) \\
    & + m (l \frac{d \beta_c}{d\lambda} + \frac{d\phi}{d\lambda})\sin(2(\beta_c l +\phi)).
  \end{split}
  \label{eq_WLD}
\end{equation}

Notably, 

\begin{equation}
  \frac{d m}{d \lambda} \propto \frac{d \Delta n_{eff}}{d \lambda} = \frac{d(n_{eff, 1} - n_{eff, 2})}{d \lambda}, 
  \label{dm_dlam}
\end{equation}
where $n_{eff, 1}$ and $n_{eff, 2}$ denote the individual effective refractive indices of the two waveguides being coupled respectively.
Further,
\begin{equation}
  \frac{d \beta_c}{d \lambda} = -\frac{\pi}{\lambda^2} \Delta n_g,
  \label{eq_betac}
\end{equation}
where $\Delta n_g  = n_{g,even}-n_{g,odd}$. $n_g$ is the group refractive index, $n_{g,*}=n_{eff,*}-\lambda \partial n_{eff,*}/\partial \lambda$ (* is even or odd).

\begin{figure*}[!t]
  \centerline{\includegraphics{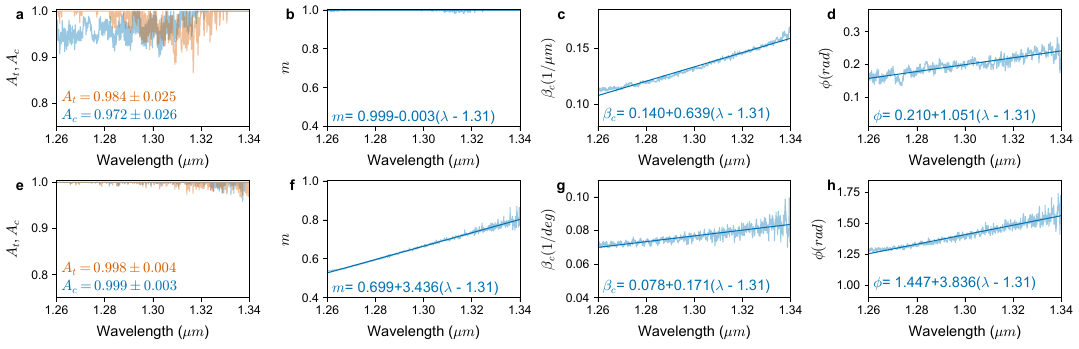}}
  \caption{Extracted model parameters wavelength response with linear fitting based on experimental data. $A_t$, $A_c$, $m$, $\beta_c$, and $\phi$ for the straight DC (a-d) and for the bent DC (e-h). The mean and standard deviation values for $A_t$ and $A_c$ for the straight (a) and bent DC (e) show the reduction of the loss by the introduction of the low-loss bends in the bent DC as compared to the traditional circular bends in the straight DC. The model parameters show expected wavelength dependence of the parameters. The symmetric straight DC has a matching coefficient, $\mathrm{m \approx 1 }$, as expected (b), while the bent DC has a wavelength dependent matching coefficient with $m = 0.699$ at $\lambda = 1.31$.  The straight DC has a coupling length of $ \mathrm{ 5.228\ \mu m}$ and a gap of  $ \mathrm{0.1\ \mu m}$.  The bent DC has a bending radius of $\mathrm{25\ \mu m}$, a gap of $\mathrm{0.1\ \mu m}$, and a coupling angle of $\mathrm{8.5\ degrees}$.
  \label{params_fitting}}
\end{figure*}


According to Eq.~\ref{eq_WLD}, and ignoring the bends coupling contribution ($\phi$) in the subsequent analysis, in order to achieve a broadband response for a directional coupler (i.e. $d\kappa^2 / d \lambda  = 0$), one would aim to achieve two criteria simultaneously, namely  $d m / d \lambda  = 0$ and $\Delta n_g = 0$.
The $dm/d \lambda = 0 $ criterion is automatically satisfied in the symmetric straight DC. This is because $n_{eff,1}=n_{eff, 2}$ for all wavelengths so that $dm/d \lambda = 0$ according to Eq.~\ref{dm_dlam}.
Nevertheless, a parameter sweep for the gaps and waveguide widths of the symmetric straight DC reveals that the $\Delta n_g = 0$ criterion is not possible, i.e. $\Delta n_g  $ is consistently positive, as depicted in Fig.~\ref{bdc_contour_plots}(a). 
This outcome underscores that the remaining second term of Eq.~\ref{eq_WLD} will only be nullified at the zeros of the sinusoid (i.e. at 0 or 1 cross coupling values). Therefore, achieving broadband behavior at an arbitrary coupling ratio with a straight symmetric DC, based on strip waveguides, is not possible. 

A similar analysis was made for the bent DC.
In order to achieve broadband coupling $\Delta n_g $ and $d \Delta n_{eff} / d \lambda $ are evaluated across various gaps and bending radii, as depicted in Fig.~\ref{bdc_contour_plots}(b) and Fig.~\ref{bdc_contour_plots}(c), respectively. 
In contrast to the straight DC, the bent DC can achieve $\Delta n_g = 0$ 
by tuning the design parameters as shown in Fig.~\ref{bdc_contour_plots}(b).
Further, the $d \Delta n_{eff} / d \lambda = 0 $ criterion is no longer automatically satisfied due to the asymmetry, in contrast to the straight symmetric DC. Only at a specific combination of gap-radius pairs that $d \Delta n_{eff} / d \lambda = 0$, as shown in Fig~\ref{bdc_contour_plots}(c).
However, $\Delta n_g = 0$ and  $d \Delta n_{eff} / d \lambda = 0$  can not be achieved with the same device parameters.
On the other hand, we observe that both $\Delta n_g $ and  $d \Delta n_{eff} / d \lambda$ can take on negative, zero, or positive values in the bent DC. This indicates the feasibility of fine tuning the design parameters in order to operate in a region where both terms of Eq.~\ref{eq_WLD} cancel each other out and eventually achieve broadband coupling, i.e.  $d\kappa^2 / d \lambda = 0$.

\begin{figure}[!b]
  \centerline{\includegraphics{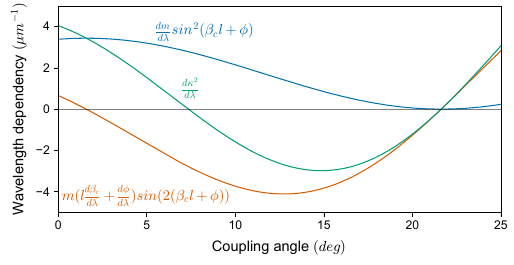}}
  \caption{Cross coupling wavelength derivative ($d\kappa^2/ d \lambda$, Eq.~\ref{eq_WLD}) of the bent DC, along with its constituent terms, demonstrating that broadband coupling (i.e. $d\kappa^2/ d \lambda = 0$ ) can take place at the intersection of the positive and negative parts of the sinusoid terms at a specific design regime. 
  The bent DC has a bending radius of $\mathrm{25\ \mu m}$ and a gap of $\mathrm{0.1\ \mu m}$.
  \label{derivative_terms}}
\end{figure}

In order to build the coupling model of the straight symmetric DC and the bent DC experimentally, Eq.~\ref{eq_for_pc} is rewritten as:

\begin{equation}
  \begin{cases}
  r^2=A_t (1-m \sin^{2}(\beta_c l+\phi)), \\
  \kappa^2=A_c m \sin^{2}(\beta_c l+\phi),
  \label{eq_for_pc_w_loss}
  \end{cases}
\end{equation}
where $A_t$ and $A_c$ are introduced to address the coupling loss of the through and cross ports respectively.
Several straight DCs and bent DCs were fabricated and measured. The fabrication is done using IMEC's most advanced iSiPP300 platform allowing high waveguide quality and access to feature dimensions well below 100 nm thanks to the 193 nm immersion lithography ~\cite{ImecSiliconPhotonics_2023a}. All the waveguides are strip-based silicon waveguides with nominal silicon thickness of 220 nm and width of 380 nm. 
In particular, the cutback method was used for the bent DC measurements for robust coupling ratio measurements, where six identical bent DCs were cascaded, and the power cross (through) coupling ratio is extracted from the slope of the linear regression between the transmitted powers (Th or Cr ports) and the port numbers as shown in Fig.~\ref{cascade_bent_dc}. 
Based on these optical measurements, the model parameters in Eq.~\ref{eq_for_pc_w_loss}, namely $A_t$,$A_c$, $m$,  $\beta_c, $ and $\phi$, were extracted by fitting the coupling ratios into the sinusoids with respect to the coupling length or angle at each measured wavelength as shown in Fig.~\ref{sin_fitting}(a) for the straight DC, and in Fig.~\ref{sin_fitting}(b) for the bent DC.
The sinusoidal fitting (depicted as a solid line) is shown to be in good agreement with the experimental data (shown in dots).
Through the fitting of the sinusoidal model as shown in Fig.~\ref{sin_fitting}, the model parameters were extracted, where we show 
a significant improvement in the excess loss in the bent DC with low-loss bends (Fig.~\ref{params_fitting}(e)) as compared to the traditional straight DC design with traditional circular bends (bending radius of 5 $\mathrm{\mu m}$, Fig.~\ref{params_fitting}(a)), where the through port transmission improved from $\mathrm{-0.070 \pm 0.109 \ dB}  $ to $\mathrm{-0.007 \pm 0.018  \ dB}$, while the cross port transmission improved from $\mathrm{- 0.126 \pm 0.115 \ dB}  $ to $\mathrm{- 0.003 \pm 0.013  \ dB} $, marking more than 40-fold decrease in the worst excess loss, thanks to the use of the low-loss bends, where the bends have both continuous curvature and curvature derivative at all connections \cite{LowLossWaveguide_2023a}. The rest of the parameters are linearly fitted with respect to the working wavelength as shown in Fig.~\ref{params_fitting}(b-d) for the straight DC and in  Fig.~\ref{params_fitting}(f-h) for the bent DC.

To the best of our knowledge, this is the lowest reported excess loss for a silicon $2 \times 2$ splitter.
This shows a substantially promising prospect for this device to be used in highly dense PICs.
Further, as expected, $m \approx 1 $ for the straight symmetric DC, while it is 0.699 at the central wavelength of 1.31 $\mathrm{\mu m}$ for the bent DC due to the asymmetry. $m$ (of the bent DC), $\beta_c$, and $\phi$ exhibit wavelength dependence that could be well represented by a linear fitting as shown in Fig.~\ref{params_fitting}(f-h). The wavelength dependence of the bent DC's parameters will be shown to cancel each other out at a specific regime in the following analysis. 

Based on the experimentally extracted parameters of the bent DC (Fig.~\ref{params_fitting}(e-h)), the two terms of the cross coupling derivative (Eq.~\ref{eq_WLD}) are plotted in Fig.~\ref{derivative_terms}. 
It is shown that a broadband response for the bent DC (i.e. $d\kappa^2 / d \lambda = 0$) is feasible at the intersection of the positive part of the first term (plotted in blue) and the negative part of the second term (plotted in orange). 
At this intersection point, the wavelength dependence of the parameters cancel each other out.
This implies the satisfaction of the broadband condition:
\begin{equation}
  \frac{d m}{d \lambda} \sin^{2}(\beta_c l+\phi) =
  - m (l \frac{d \beta_c}{d\lambda} + \frac{d\phi}{d\lambda})\sin(2(\beta_c l +\phi))
  \label{broadband_condition}
\end{equation}

\section{Broadband, ultra low-loss, and Tolerant 0.5:0.5 coupler with 300 mm wafer mapping}

\begin{figure}[!t]
  \centerline{\includegraphics{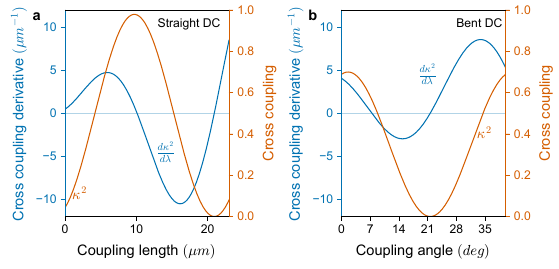}}
  \caption{The cross coupling $(\kappa^2)$ and the cross coupling derivative $(d\kappa^2/ d \lambda)$ as a function of the coupling length for the straight DC (a) and as a function of the coupling angle for the bent DC (b). The introduced asymmetry presented in the bent DC shifted the broadband cross coupling point ($\kappa^2 \vert_{d \kappa^2 / d \lambda = 0 }$) from $\kappa^2 \vert_{d \kappa^2 / d \lambda = 0 } = 1 $ to $\kappa^2 \vert_{d \kappa^2 / d \lambda = 0 } \sim 0.5 $, this idea makes it possible to have broadband coupling at distinct coupling ratios based on the bending radius value. 
  All the designs have a gap of $\mathrm{0.1\ \mu m}$. 
  The bent DC has a bending radius of $\mathrm{25\ \mu m}$ and a gap of $\mathrm{0.1\ \mu m}$.
  \label{cross_coup_and_deriv}}
\end{figure}
\begin{figure}[!t]
  \centerline{\includegraphics{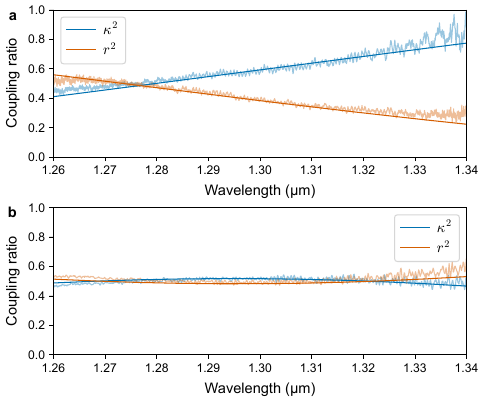}}
  \caption{The measured coupling ratios of the traditional straight DC (a) and the proposed bent DC (b) at $0.5:0.5$ coupling. 
  The straight DC has a  coupling length of 5.228 $\mathrm{\mu m}$ and a gap of 0.1 $\mathrm{\mu m}$.
  The bent DC has a bending radius of 25 $\mathrm{\mu m}$, a gap of 0.1 $\mathrm{\mu m}$, and a coupling angle of $\mathrm{8.5\ degrees}$.
  The bent DC shows broadband coupling with a minimal variation of 0.051 achieving 7.67 times reduction in coupling variation as compared to the straight DC. Non transparent lines show the model fitting, while the transparent lines show the measured experimental data.
    \label{3db_comparison}}
\end{figure}

\begin{figure}[!t]
  \centerline{\includegraphics{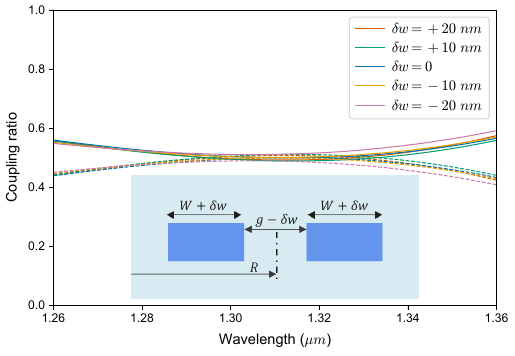}}
  \caption{Fabrication tolerance of the proposed $0.5:0.5$ bent DC. Coupling spectrum simulation results of the nominal design ($\delta w = 0$) along with waveguide width deviations of -20, -10, 10, 20 nm. Through couplings are shown in solid lines and the cross couplings are shown in dashed lines.  
  The nominal design has a bending radius ($R$) of $\mathrm{25\ \mu m}$, gap ($g$) of $0.1\ \mu m$, and coupling angle of $\mathrm{8.5\ degrees}$. The inset figure depicts the waveguide width ($W$) deviation and the corresponding gap deviation for the proposed SOI-based bent DC. 
  \label{fab_tol}}
\end{figure}

\begin{figure*}[!t]
  \centerline{\includegraphics{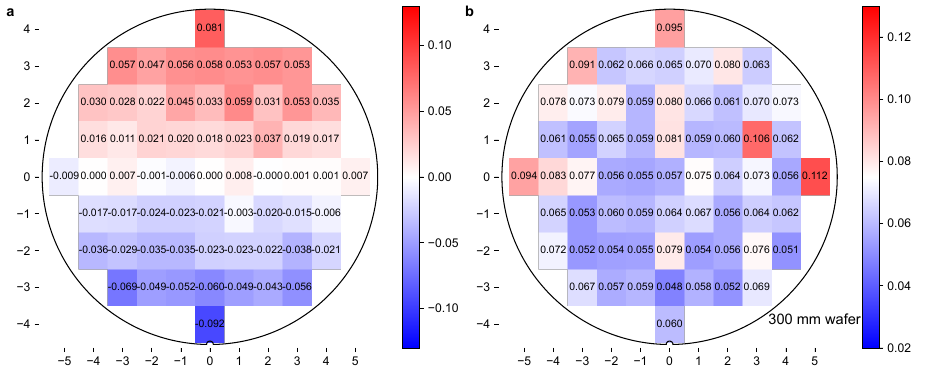}}
  \caption{ Average cross coupling deviation with respect to the central die (a) and cross coupling variation (b) over 80 nm bandwidth for the proposed 0.5:0.5 bent DC splitter over the 300 mm wafer, covering all 63 dies. Low variation is shown over all dies except a relatively higher variation at the edges of the wafer, as expected due to process variation. The bent DC has a bending radius of $\mathrm{25\ \mu m}$, a gap of $\mathrm{0.1\ \mu m}$, and a coupling angle of $\mathrm{8.5\ degrees}$.
  \label{wafer_mapping}}
\end{figure*}

To showcase the broadband coupling and further investigate the impact of asymmetry on the coupling wavelength dependence for the bent DC, we present the cross coupling along with its derivative for the fabricated 0.5:0.5 splitters using straight and bent DC with varying lengths or coupling angles, as depicted in Fig.~\ref{cross_coup_and_deriv}(a) and (b), respectively.
In the case of the straight DC (Fig.~\ref{cross_coup_and_deriv}(a)), $d\kappa^2 / d\lambda = 0$ only at the valley and the peak of the sinusoid (i.e. at 0 and 1 cross coupling ratios, respectively). 
Further, at the 0.5:0.5 point, $d\kappa^2 / d\lambda \approx 4~\mu m^{-1}$, very close to the peak of the derivative, showing high wavelength dependence.
Conversely, in the case of the bent DC (Fig.~\ref{cross_coup_and_deriv}(b)), $d\kappa^2 / d\lambda = 0$ at the cross coupling value of $\sim 0.5$.

The coupling length and the coupling angle corresponding to 0.5 coupling ratio were extracted from Fig.~\ref{cross_coup_and_deriv} for both the straight and bent DC, respectively. Further, their coupling spectrum is depicted in  Fig.~\ref{3db_comparison}(a)  for the straight DC and in Fig.~\ref{3db_comparison}(b) for the bent DC, showing consistency with the presented analysis.
In particular, in Fig.~\ref{3db_comparison}(a), the straight DC demonstrates high wavelength dependence, recording 0.391 coupling variation over 80 nm wavelength range.
In contrast, the bent DC (Fig.~\ref{3db_comparison}(b)) exhibits broadband coupling and showcased a minimal cross coupling variation of 0.051.
Quantitatively, the coupling variation of the proposed bent DC is 7.67 times lower than that of the straight DC.
For the bent DC, around $\lambda = 1.3\ \mathrm{\mu m}$, $\kappa^2 \approx 0.5$ where $d \kappa^2 /d \lambda$ is substantially reduced to zero enabling the broadband behavior of the coupler, in contrast to a highly variant coupling behavior of the straight DC. 
Furthermore, the model presented in Fig.~\ref{params_fitting} for the straight and bent DC was used to fit the experimental coupling data, as shown in Fig.~\ref{3db_comparison} by solid curves with deeper colors, showing agreement with the experimental data.

To assess the fabrication tolerance of the proposed $0.5:0.5$ bent DC, the coupling behavior of devices with waveguide width deviations ($\delta w$) of $\mathrm{-20,\ -10,\ 0,\ 10,\ 20\ nm}$ are simulated using 3D-FDTD model as shown in Fig.~\ref{fab_tol}. 
It is shown that the device still performs well even with $\mathrm{\delta w = \pm 20 \ nm}$, showcasing the tolerance of the proposed device. 
In particular, the cross coupling variation over 80 nm wavelength range away from 0.5 is 0.058 for $\mathrm{\delta w = \pm 10 \ nm}$, while it is 0.061  for $\mathrm{\delta w = \pm 20\ nm}$.

Futhermore, the robustness of the proposed design across the 300 mm wafer is also investigated.  
Complete wafer measurements were conducted for two metrics, namely the cross coupling deviation, averaged over 80 nm wavelength range, with respect to the central die (Fig.~\ref{wafer_mapping}(a)) and the cross coupling variation (Fig.~\ref{wafer_mapping}(b)) over 80 nm wavelength range, covering all 63 dies. 
Around the nine central dies, the average cross coupling deviation from the central die ranged between 0.003 and 0.023, where the cross coupling variation ranged between 0.055 and 0.081.
Overall, the die with the largest coupling variation was die (5, 0) at the extreme edge of the wafer where process variations are usually high. 
Die (5,0) has 0.112 maximum cross coupling variation over 80 nm wavelength range, with an average cross coupling deviation of 0.007.
In general, the results indicate low variations across most dies, with a relatively high value observed only at the extreme edges of the wafer.
This further illustrates the potential of the large-scale use of the proposed device.

\section{Broadband coupling with arbitrary coupling ratios: model and coupling examples}

\begin{figure}[!t]
  \centerline{\includegraphics{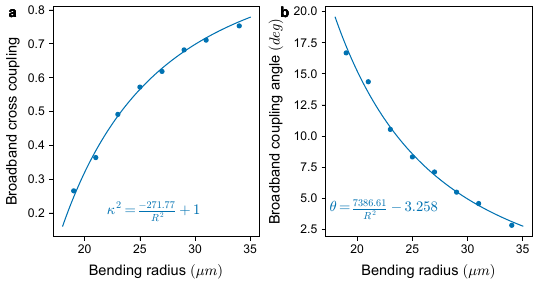}}
  \caption{The proposed model for extracting broadband bent DCs with arbitrary coupling ratios. 
  Broadband cross coupling values as a function of the bending radius (a). The corresponding coupling angle as a function of bending radius (b). The experimental data is shown in dots while the fitting model is depicted as a solid line. The coupling gap and waveguides widths are fixed as $\mathrm{0.1\ \mu m}$ and $\mathrm{0.38\ \mu m}$, respectively.
  \label{model}}
\end{figure}

The asymmetry is inversely proportional to the bending radius ($R$), with higher asymmetry, higher phase mismatch takes place and the maximum coupling ratio goes down. A comprehensive fitting was conducted for broadband coupling ratios, considering both bending radius and coupling angle while maintaining a fixed gap of $\mathrm{0.1\ \mu m}$ and waveguide widths of $\mathrm{0.38\ \mu m}$, as shown in Fig.~\ref{model}.
As depicted in Fig.~\ref{model}(a), with the increase of the bending radius, the mismatch decreases and higher cross coupling values are achieved.
Additionally, the increase of the bending radius (i.e. decrease in the mismatch) implies a decrease in the required coupling angle to achieve a specific coupling value, as illustrated in Fig.~\ref{model}(b).
This fitting provides a valuable tool for achieving broadband  coupling for arbitrary coupling ratios.

\begin{table*}[t!]
  \centering
  \caption{Performance comparison of fabricated $2 \times 2 $ splitters.}
  \begin{tabular}{cccccc}
    \hline
      Reference & Structure & 0.5:0.5 coupling variation & Excess loss & Device length & Coupling ratios  \\
      \ &   & (over 80 nm) & (dB) &  ($\mathrm{\mu m}$) & demonstrated \\ 
      \hline
      \cite{BroadbandAngledArbitrary_JOLT2020a} & MMI & $>0.037$ & $<1.3$ & 75 & \makecell{0.07, 0.15, 0.3, 0.4, \\ 0.5, 0.6, 0.7, 0.8, 0.9} \\ 
      \cite{UltraBroadbandChip_OE2024a} & Rib-waveguide-based ADC & 0.082 & $< 0.22 $ & 79 & 0.5, 0.7, 0.9   \\ 
      \cite{CompactUltraBroadband_IPTL2024a}& Rib-waveguide-based ADC &  $ 0.038 $ & $<0.18$ & 108 & 0.5 \\  
      \cite{AdiabaticityEngineeredSilicon_IPJ2023a}& ADC &  $>0.1$   & - & 67 & 0.5 \\  
      \cite{BroadbandSiliconPhotonic_OE2015a} & MZI &   0.11 & $<1$  & 31.4 & 0.1, 0.2, 0.3, 0.4, 0.5 \\
      \cite{FabricationTolerantWavelength_OE2022a} & MZI circuit & 0.051 & $<0.38$ & 60 & 0.05, 0.2, 0.3, 0.5 \\ 
      \cite{UltraBroadband2_OL2018a} & SWG-assisted ADC & 0.039 &  $<0.11$ & 35 & 0.5 \\ %
      \cite{BroadbandSiliconInsulator_SR2017a} & Bent DC & 0.18  &  - & $>20$ & 0.5  \\ 
      \cite{PolarizationInsensitiveBroadband_OL2017a} & Bent DC & 0.13  & $<1$ & 50 & 0.5   \\ 
      \cite{HighPerformance2×2_JOLT2023a}& Semi-inverse designed bent DC &  0.106 &  $<0.05$ & 28.8 & 0.5 \\ 
      This work & Bent DC & 0.051 & $\mathrm{0.003 \pm 0.013}$ & 27.5 & \makecell{0.4, 0.5, 0.6, 0.7 with \\ full coupling model} \\ 
      \hline
  \end{tabular}
  \label{tab_splitter_comparison}
\end{table*}

\begin{figure}[t!]
  \centerline{\includegraphics{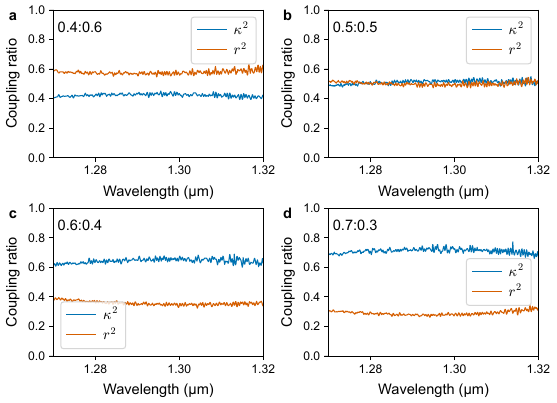}}
  \caption{Examples of broadband bent DCs with arbitrary coupling ratios in accordance with the proposed model, where the through and cross coupling are shown as a function of wavelength. Devices with broadband 0.4 (a), 0.5 (b), 0.6 (c), and 0.7 (d) cross coupling values are presented. 
  The bending radii and the coupling angles of the devices are: $\mathrm{23\ \mu m}$ and $\mathrm{12\ degrees}$ (a), $\mathrm{25\ \mu m}$ and $\mathrm{8.5\ degrees}$ (b), $\mathrm{29\ \mu m}$ and $\mathrm{6\ degrees}$ (c), and $\mathrm{34\ \mu m}$ and $\mathrm{4\ degrees}$ (d).
  All the designs have a gap of $\mathrm{0.1\ \mu m}$. 
  \label{examples}}
\end{figure}

The fitting depicted in Fig.~\ref{model} is subsequently  employed to extract broadband splitters with arbitrary coupling ratios.
As illustrated in Fig.~\ref{examples}, multiple examples of devices with broadband cross coupling ratios of 0.4 (a), 0.5 (b), 0.6 (c), and 0.7 (d) are shown. 
The corresponding cross coupling variations over 50 nm wavelength range for these ratios are minimally equal to 0.023, 0.023, 0.038, and 0.034, respectively. 
This affirms the potential of the proposed methodology for designing broadband coupling ratios with flexibility.

Compared to the results existing in literature (Tab.~\ref{tab_splitter_comparison}),  our proposed splitter stands out as a high-performance design that simultaneously meets all essential criteria of low wavelength dependence, ultra low-loss coupling, compact footprint, support for arbitrary coupling ratios, and high fabrication tolerance.
Specifically, we were able to introduce the first experimental-based model to achieve broadband coupling with arbitrary coupling ratios using the bent DC.
On that basis, we present a compact $2 \times 2$ splitter with low coupling variation and the lowest reported excess loss for a $2\times2$ splitter to our best knowledge.
Furthermore, our proposed device exhibits consistently low variations across the 300 mm wafer, indicating its fabrication tolerance. 
We augmented the wafer measurements with waveguides width variations study, where the proposed device showed a cross coupling variation away from 0.5 of 0.061 over the proposed wavelength range for $\mathrm{\delta w = \pm 20\ nm}$.
These features highlight the potential of our proposed device to be used in mass production reliably.

\section{Conclusion}

In conclusion, this study presented a comprehensive CMT-based and experimental analysis for a bent DC with low wavelength sensitivity.  
Leveraging this analysis, a model was derived in order to achieve broadband bent DCs with arbitrary coupling ratios.
In particular, we used the proposed model to verify the broadband bent DCs at a few coupling ratios including 0.4, 0.5, 0.6, and 0.7.
As a benchmark, we demonstrated a  0.5:0.5 splitter showcasing the least coupling variation in a bent DC to our best knowledge with 0.051 cross coupling variation over 80 nm wavelength range. Additionally, the proposed 0.5:0.5 splitter exhibited the least excess loss in the literature to our best knowledge ($\mathrm{0.003 \pm 0.013 \ dB}$), along with the capability of achieving arbitrary coupling ratios, compactness (with a length of $\mathrm{27.5\ \mu m }$), and high fabrication tolerance (with a cross coupling variation of 0.061 with $\mathrm{\delta w = \pm 20 \ nm}$ over 80 nm wavelength range). 
Wafer mapping analysis confirmed the robustness of the proposed device for practical applications, with low variation observed over a 300 mm wafer, exhibiting a maximum coupling variation value of 0.112 and a corresponding average cross coupling deviation of 0.007 at the extreme edge of the wafer over 80 nm wavelength range. Overall, the proposed bent DCs meet all the aforementioned essential requirements and exhibit reliability for integration into highly dense PICs.


\bibliographystyle{IEEEtran}
\bibliography{reference}

\end{document}